\documentclass[superscriptaddress,aps,prl,preprintnumbers,nofootinbib,preprint,11pt]{revtex4}
\usepackage[utf8]{inputenc}
\usepackage[citecolor=blue,urlcolor=blue,unicode=true,pdfusetitle,
bookmarks=true,bookmarksnumbered=true,bookmarksopen=true,bookmarksopenlevel=3,
breaklinks=false,pdfborder={0 0 1},backref=false,colorlinks=true]{hyperref}
\usepackage{bm}
\usepackage{latexsym}
\usepackage{dcolumn,xcolor}
\usepackage[normalem]{ulem}
\usepackage{amsmath,amsfonts,amssymb}
\usepackage{graphicx,epsfig}
\usepackage{soul}
\usepackage{ulem}
\usepackage{amsthm}
\usepackage{color,float} 

\usepackage{amssymb}
\usepackage{braket}
\usepackage{physics}
\usepackage{tikz}

\begin{document}
\title{Can quantum gravity be both consistent and complete?}

\author{Mir Faizal\footnote{Email: \href{mailto:mirfaizalmir@gmail.com}{mirfaizalmir@gmail.com}}}
\affiliation{ \scriptsize{Irving K. Barber School of Arts and Sciences, 
  University of British Columbia - Okanagan, Kelowna,
British Columbia V1V 1V7, Canada.}}
\affiliation{\scriptsize{Canadian Quantum Research Center, 204-3002 32 Ave, Vernon, BC V1T 2L7, Canada.}}
\affiliation{ \scriptsize{Department of Mathematical Sciences, Durham University, Upper Mountjoy, Stockton Road, Durham DH1 3LE, UK.}}
\affiliation{\scriptsize{Faculty of Sciences, Hasselt University, Agoralaan Gebouw D, Diepenbeek, 3590, Belgium.}}
\author{Lawrence M. Krauss\footnote{Email: \href{mailto:lawrence@originsproject.org}{lawrence@originsproject.org}}}
\affiliation{\scriptsize{Origin Project Foundation, Phoenix, AZ 85018, USA. (Corresponding author)}}
\author{Arshid Shabir\footnote{Email: \href{mailto:aslone186@gmail.com}{aslone186@gmail.com}}}
\affiliation{\scriptsize{Canadian Quantum Research Center, 204-3002 32 Ave, Vernon, BC V1T 2L7, Canada.}}
\author{Francesco Marino\footnote{Email: \href{mailto:francesco.marino@ino.cnr.it}{francesco.marino@ino.cnr.it}}}
\affiliation{\scriptsize{CNR-Istituto Nazionale di Ottica and INFN, Via Sansone 1, I-50019 Sesto Fiorentino (FI), Italy.}}
\author{Behnam Pourhassan\footnote{Email: \href{mailto:b.pourhassan@du.ac.ir}{b.pourhassan@du.ac.ir}}}
\affiliation{\scriptsize{School of Physics, Damghan University, Damghan 3671645667, Iran.}}
\affiliation{\scriptsize{Center for Theoretical Physics, Khazar University, 41 Mehseti Street, Baku, AZ1096, Azerbaijan.}}

\date{\today}

\begin{abstract}
General relativity, despite its profound successes, fails as a complete theory due to presence of singularities.  While it is widely believed that quantum gravity has the potential to be a complete theory, in which spacetime consistently emerges from quantum degrees of freedom through computational algorithms, we argue that this goal could be fundamentally unattainable. We examine how this limitation could emerge in various contexts, depending on whether or not every mathematically valid result is physically realized. In the first case, G\"odel’s incompleteness theorems, along with related results by Tarski and Chaitin, imply that no theory formulated as a formal axiomatic system can be complete, and that within any computational framework, a fully consistent internal truth predicate is impossible. In the second case, if only a subset of mathematical truths is realized in nature, we argue that this selection cannot be determined by any purely computational process. Hence, a meta-theoretical approach based on non-algorithmic understanding is indispensable in every case. We discuss some possible consequences of this observation for describing physical systems and note that a non-algorithmic approach should be essential for any theory of everything.  \\

\textbf{Essay written for the Gravity Research Foundation 2025 Awards for Essays on Gravitation.}
 \end{abstract}

\maketitle

\section{Introduction}
Physics was developed to describe and predict phenomena occurring in spacetime. However, the theory of general relativity--linking gravity to the curvature of spacetime--is a theory of spacetime itself. It has had  overwhelming success in explaining gravitational phenomena, from the perihelion precession of Mercury to the recent detection of gravitational waves \cite{Einstein1915, Abbott2016}. Despite its success, general relativity predicts its own breakdown  at singularities, which occur at the center of black holes, and at the initial instant of the big bang \cite{Penrose1965, Hawking1970}. At these points the very structure of spacetime breaks down and becomes ill-defined.  Thus, this theory is not complete as it cannot describe physics at such points.

Singularities are not exclusive to general relativity; they emerge across various fields of physics \cite{Arnold92,Berry2023}, at scales where a given model no longer adequately describes the system. They are expected to disappear when the full theoretical framework is applied. A prominent example is flow discontinuities in classical fluid mechanics, which are associated to curvature singularities of an effective (acoustic) metric and are resolved through a full quantum hydrodynamic framework  \cite{Faccio2016,Braunstein:2023jpo}. Similarly, in different approaches to quantum gravity, such as loop quantum gravity (LQG) and string theory, curvature singularities are replaced with finite, well-defined structures. In loop quantum cosmology (LQC), a symmetry-reduced model of LQG, the big bang singularity is resolved through a "quantum bounce," where the universe transitions from contraction to expansion \cite{Bojowald2001, Ashtekar2006}. In string theory, the fuzzball paradigm replaces black hole singularities with extended structures  \cite{Mathur2005, Mathur2008}.   

The disappearance of singularities suggests something deeper: spacetime as represented by the smooth manifold of classical relativity is an emergent structure that breaks down in certain situations. This is similar to the emergence of effective manifolds in condensed matter physics from a mean-field representation of underlying quantum dynamics \cite{Braunstein:2023jpo}.
In the context of string theory, using double field theory (DFT),  T-duality can be incorporated as a fundamental symmetry \cite{Hohm2010}.  T-folds, a  concept based on it, extend spacetime geometry by allowing transition functions involving T-duality transformations. This further reinforces the idea that spacetime is an emergent structure within quantum gravity, as spacetime is  not always  well-defined in this case \cite{Hull2005}. Likewise, spin foam models in loop quantum gravity (LQG) indicates the emergent nature of  spacetime. These models describe spacetime as arising from a discrete network of quantum structures \cite{Perez2013}. Thus, both string theory and LQG exemplify the emergent nature of spacetime, wherein the smooth manifold of classical relativity arises from a deeper quantum gravitational degrees of freedom.

These developments strongly align with Wheeler's "it from bit" proposal, along with its contemporary formulations within string theory \cite{Jafferis:2022crx, VanRaamsdonk:2020ydg} and loop quantum gravity (LQG) \cite{Makela:2019vgf}, which assert that information fundamentally underpins physical reality \cite{Wheeler1990}.
The inability of classical theories to describe singularities further supports this perspective: spacetime, as an emergent phenomenon, cannot account for regions where its underlying quantum information structure cannot be described in terms of spacetime geometry. 
While `it'--representing spacetime and matter fields in it--is not complete, it is expected that `bit'--representing quantum gravity--could be complete and consistent computational `theory of everything'. 
We will argue here, however, that it is not possible to construct such a theory based on algorithmic computations. 

{ Here we clarify two distinct philosophical attitudes toward the relation between mathematics, the theory of everything, and physical reality. The first approach can be referred to as Universal Actualism.  
  In this first stance, championed by Weinberg   \cite{weinberg2011dreams}, every mathematically sound deduction from the theory of everything is, in principle, physically actualized.  The theory of everything is identified with a maximally expressive set of axioms; once the axioms are fixed, all their theorems are regarded as concrete physical facts about the cosmos.  Existence is therefore equated with formal derivability, and the distinction between “possible” and “actual” collapses into pure mathematics.  Recent proponents of this view include “Mathematical Universe Hypothesis,” which similarly equates physical existence with membership in the set of consistent mathematical structures \cite{tegmark2008mu}.

 The second philosophical position may be referred to as Selective Actualism.
 In this approach proposed by Schmidhuber, it is argued that while all computable universes exist as valid abstract programs, only a select subset are physically actualized \cite{schmidhuber1997computer}. A clear distinction is made between what is a mathematically valid deduction from the theory of everything and what is physically realized—the latter being only a subset of the former. This selection is not determined by computation alone, but rather by a meta-mathematical principle that relies on non-algorithmic understanding.

Although Selective Actualism clearly requires the existence of non-algorithmic understanding, the purpose of this essay is to argue that non-algorithmic understanding should be also essential even within the framework of Universal Actualism. G\"odel’s incompleteness theorems, together with related results by Tarski and Chaitin, suggest that no theory expressed as a formal axiomatic system can be both complete and consistent. This suggests that a form of non algorithmic understanding is required, even in a context where all mathematically valid results are expected to be physically realized.
}

\section{Quantum Gravity} 
Quantum gravity aims to unify the principles of quantum mechanics and general relativity within a single consistent framework.
 {Despite significant advancements in approaches such as string theory \cite{Green1987, Polchinski1998} and loop quantum gravity (LQG) \cite{Rovelli2004, Thiemann2007}, a complete and definitive theory remains elusive. So, we have different approaches to quantum gravity, and based on those different approaches multiple axiomatic constructions of quantum gravity have been proposed \cite{Witten:1985cc, Ziaeepour:2021ubo, Faizal2024, bombelli1987spacetime, Majid:2017bul, DAriano:2016njq, Arsiwalla:2021eao}. }
 In general, we first note that any viable candidate for quantum gravity must adhere to fundamental axioms that ensure both physical and mathematical coherence.
Furthermore, such a formal mathematical axiomatic system, $\mathcal{F}_{QG}$, generates spacetime rather than existing within it \cite{Faizal2023}. 
These include both physically motivated  axioms of quantum gravity denoted by $\mathcal{F}_{PQG}$, and the axioms of the mathematics quantum gravity is built upon (such as Zermelo-Fraenkel set theory with the Axiom of Choice) denoted by $\mathcal{F}_{MQG}$. Here, we have $\mathcal{F}_{QG} = \mathcal{F}_{PQG} \cup \mathcal{F}_{MQG}$.

The essential axioms required for quantum gravity to qualify as a theory of everything, can  be summarized as follows:

{1. Effectively Axiomatizable}: The axioms $\{A_1, A_2, \dots\}$ of $\mathcal{F}_{QG}$ are finite. This is important for the theory to be computationally well-defined.  Spacetime should emerge as a consequence of these axioms \cite{Faizal2023, Braunstein:2023jpo}.
 For example, in string theory, spacetime emerges from the dynamics of extended objects \cite{Seiberg2006, Polchinski1998}, or as a result of quantum information using holography \cite{Jafferis:2022crx, VanRaamsdonk:2020ydg}, and in LQG, spacetime is an emergent feature of spin networks and spin foams \cite{Perez2013, Rovelli2004}, and can be again derived from quantum information \cite{Makela:2019vgf}. 

{2. Sufficient Expressiveness}: The system $\mathcal{F}_{QG}$ is capable of encoding basic arithmetic operations over natural numbers, including addition and multiplication. Quantum gravity must describe physical phenomena, which reduce to both classical spacetime and standard quantum mechanics in certain limits, and so it must be capable of allowing numerical calculations (e.g., scattering amplitudes, curvature, entropy etc), which inherently require arithmetic. It maybe noted that both string theory and LQG reproduce results from general relativity and quantum mechanics/quantum field theory in  appropriate limits \cite{Polchinski1998, Green1987, Rovelli2004, Thiemann2007}. 

 {3. Consistency}:
A theory of quantum gravity must be internally consistent, and so $\mathcal{F}_{QG}$ does not derive any contradictions.  Apart from mathematical consistency, it should be physical consistent, avoiding unphysical results like negative probabilities. String theory achieves consistency through anomaly cancellation \cite{Green1984, Polchinski1998}, while LQG ensures it through the rigorous quantization of geometric degrees of freedom \cite{Ashtekar1986, Rovelli2004}. 

 {4. Completeness \footnote{In this context, completeness should be understood as a complete description of all physical phenomena. 
It should not be confused with mathematical completeness, which is considered here only in the context of Universal Actualism. In Selective Actualism, completeness means that everything physical should be a computational or G\"odelian consequence of the theory, although not every mathematical consequence of the theory is necessarily physical.}:
The theory must describe phenomena across all scales, from the Planck scale to cosmological scales.  This includes the resolution of singularities, such as those found in black holes and the big bang \cite{Bojowald2001, Mathur2005}. A complete theory should also encompass the origin of universe, which is attempted via  quantum cosmology in LQG \cite{Ashtekar2006, Rovelli2004} or colliding branes in string theory \cite{Khoury:2001wf}. It should be consistent with all observational data.

These axioms should produce a formal structure that satisfies some  important  properties. 
A viable theory of quantum gravity must reduce to standard theories, in the limit those theories have been tested and verified. For example,  both string theory and LQG fulfill this requirement by recovering quantum field theories and general relativity in appropriate limits \cite{Green1987, Rovelli2004, Thiemann2007}. 

It must also properly address the measurement problem by providing a mechanism for the apparent collapse of the wavefunction.
Furthermore, this has to be intrinsic to explain the quantum-to-classical transition in cosmology \cite{Gaona-Reyes:2024qcc}. Models based on gravitationally induced objective collapse \cite{Penrose1996, Diosi1987}, suggest that quantum gravity may naturally resolve this issue by linking wavefunction collapse to spacetime structure.  This is neither random nor algorithmic nor computational, but is fundamentally related to non-algorithmic understanding.  

The theory must also explain the emergence of spacetime as a macroscopic phenomenon arising from more fundamental quantum structures.  It is important that the emergent spacetime does not lead to contradictions, such as closed timelike curves, which would require principles like the Novikov self-consistency principle \cite{novikov1989}, again based on non-algorithmic understanding.

Furthermore, spacetime singularities should be naturally resolved in quantum gravity.  Such resolution of singularities occurs  in principle in string theory due to extended objects \cite{Mathur2005, Seiberg2006}, and in   LQG due to discrete quantum geometries \cite{Bojowald2001, Ashtekar2006}. However, it is important to note that these involve Planck scale physics, and it has been demonstrated that quantum gravity prevent its own measurement at such scales \cite{Pourhassan:2023jbs, Pourhassan2025}. So, dealing with this  require non-algorithmic understanding as that described by the Lucas-Penrose argument \cite{lucas1961minds, Penrose2011, Penrose1990-PENTNM, hameroff2014consciousness, lucas_penrose_2023}. As this non-algorithmic understanding operates at a fundamental level in nature,  it is not possible to explain all natural phenomenon using computational algorithms alone. 
As such there is a more fundamental reason why quantum gravity, when based solely on computation, cannot be both consistent and complete.   

\section{G\"odel's and Tarski's Theorems in Quantum Gravity}
  We will begin by distinguishing between the computational consequences and the G\"odelian consequences of an axiomatic system describing quantum gravity \cite{Godel1931, Smith2007}. In any sufficiently expressive, consistent, and effectively axiomatizable formulation of quantum gravity, it is possible to derive certain consequences through computational algorithms. However, these computational outcomes do not encompass the full set of consequences that can be derived from the system.
  
  To begin, we consider a formal system for quantum gravity, which we denote as $\mathcal{F}_{QG}$. This system is assumed to be consistent, meaning that it does not derive any contradictions, but it is also sufficiently expressive and axiomatizable for these theorems to hold. 
G\"odel’s incompleteness theorems state that in any formal system that is sufficiently expressive, there exist mathematical statements that are true within the system but cannot be proven by the system itself \cite{Godel1931, Smith2007}. In the context of quantum gravity, as the formal system $\mathcal{F}_{QG}$ is consistent and sufficiently expressive, there must be certain mathematical statements, like $G$, which cannot be proved within the system. Such statements are true by virtue of the structure of the system but are not provable using its axioms and rules.  

So, certainly, $\mathcal{F}_{QG}$ cannot be both consistent and complete, and this is a direct implication of G\"odel's theorems. The truths that will be true will not be limited to the computational consequences of $\mathcal{F}_{QG}$, but will also include G\"odelian consequences of $\mathcal{F}_{QG}$. In Universal Actualism, these G\"odelian consequences will definitely correspond to some physical aspects of the theory. Thus, undefinability will be an intrinsic feature of Universal Actualism, and we will require a meta-theoretical approach to address them. On the other hand, in Selective Actualism, two possibilities arise: either some G\"odelian consequences are part of the physically important consequences, or none of the G\"odelian consequences are expressed as physically important consequences. 
In the first case, we still have some G\"odelian consequences, which are a part of the physical theory and hence need a meta-theoretical approach. The other case requires the identification and elimination of all G\"odelian consequences of the theory, and this selection process cannot be algorithmic. So, as this elimination of all the G\"odelian consequences is, in itself, non-algorithmic and beyond computation, it again requires a meta-theoretical approach.
In any case, we do require a meta-theoretical approach based on a non-algorithmic understanding to construct  quantum gravity \cite{Faizal2024, Faizal20240}. 

Additionally, while the system $\mathcal{F}_{QG}$ can be assumed to be consistent, it cannot prove its own consistency. This is a direct consequence of G\"odel’s second incompleteness theorem, which asserts that no sufficiently powerful formal system can prove its own consistency. Therefore, $\mathcal{F}_{QG}$ cannot prove the  {mathematical} statement $\text{Con}(\mathcal{F}_{QG})$, which asserts that $\mathcal{F}_{QG}$ is free from contradictions. If $\mathcal{F}_{QG}$ were able to prove its own consistency, it would also be able to prove the mathematical statement $G$, which contradicts the undecidability of $G$. Consequently, the consistency of $\mathcal{F}_{QG}$ must be established externally, relying on a meta-theoretical framework that transcends the formal system itself \cite{Faizal2024, Faizal20240}.  This holds true for both Selective Actualism and Universal Actualism. In Selective Actualism, this is trivially true, as the very definition of the approach does not imply that a fully complete and consistent physical theory can be derived solely from computations based on the axioms of quantum gravity. In Universal Actualism, it holds true due to G\"odel's incompleteness theorems.

We have not yet discussed whether these consequences will have important physical implications or will be trivially true due to the very structure of $\mathcal{F}_{QG} = \mathcal{F}_{PQG} \cup \mathcal{F}_{MQG}$. 
Here, we would like to note that, on physical grounds, some G\"odelian statements of $\mathcal{F}_{QG}$ should have important physical consequences.   
Different approaches to quantum gravity are based on the quantization of some underlying degrees of freedom and thus obey quantum logic. For example, string theory is based on the quantum theory of strings \cite{Green1984, Polchinski1998}, while LQG is based on a quantum theory of certain discrete degrees of freedom \cite{Ashtekar1986, Rovelli2004}.  Hence, all approaches to quantum gravity obey quantum logic.
However, quantum logic has been demonstrated to be fundamentally undecidable and G\"odelian  \cite{Fritz2021, Eisert2012, vandenNest2008measurement, lloyd1993quantum}. Even when quantum mechanics has been proposed to be modified by quantum gravity, such as in gravitational collapse models, it has been suggested that the objective collapse in the measurement problem of quantum mechanics itself is related to G\"odel's incompleteness theorem \cite{Penrose1996, Diosi1987}. Thus, there are fundamental physical consequences of every quantum system that cannot be fully captured by any algorithmic computation.
Certainly, these would be present in any formalism of quantum gravity, which would be based on some quantum logic. More directly, aligning with Wheeler’s ‘it from bit’ \cite{Wheeler1990}—along with its contemporary formulations within string theory \cite{Jafferis:2022crx, VanRaamsdonk:2020ydg} and LQG \cite{Makela:2019vgf}—we observe that quantum information in quantum gravity constitutes the basis of observable physical phenomena.
 As quantum logic for this quantum information describing quantum gravity is fundamentally undecidable and G\"odelian in nature, certain physical phenomena that follow from the quantum logic of quantum gravity must likewise be undecidable and G\"odelian.  So, as ‘bit’—or more precisely, ‘qubit’—is certainly G\"odelian \cite{Fritz2021, Eisert2012, vandenNest2008measurement, lloyd1993quantum}, ‘it’ must be G\"odelian as well.
 
Furthermore, it has been argued that even the selection of observables in general relativity is G\"odelian and thus undecidable \cite{Panagiotopoulos2023}. It seems very unlikely that quantum gravitational arguments could make it decidable.  In fact, we would possibly calculate them as expectation values of some underlying degrees of freedom. The details of the calculation will depend on the specific approach taken. However, those calculations would be undecidable and G\"odelian, as we already know their expectation value has to coincide with the results in general relativity due to the Ehrenfest's theorem. 

We now note that, within the framework of Universal Actualism, it has been argued using Weinberg's approach that if we had a final computational theory of everything, a complete theory of quantum gravity, then all scientific mathematical truths would be computationally derived from the axioms of that theory 
\cite{weinberg2011dreams}. Tarski's undefinability theorem shows that this is impossible \cite{Tarski1933, Tarski1983}.  Thus, we  now turn to the implications of Tarski's undefinability theorem for $\mathcal{F}_{QG}$. According to Tarski's theorem, the truth predicate for mathematical statements within $\mathcal{F}_{QG}$ cannot be defined within the system itself. If we assume that the truth predicate, $\text{True}_{\mathcal{F}_{QG}}(x)$, can be defined within $\mathcal{F}_{QG}$, we can construct a self-referential mathematical statement $\psi$ that asserts, "The statement $\psi$ is not true." This leads to a paradox, as $\psi$ would be true if and only if it is not true, thus violating the principle of non-contradiction. This contradiction shows that the truth predicate cannot be defined within the formal system.

The undefinability of the truth predicate has profound implications for the understanding of quantum gravity. It suggests that certain truths within the system cannot be fully understood within the system itself and require an external, meta-theoretical framework for their interpretation. This meta-theoretical framework is based on an external truth predicate, which is constructed using non-algorithmic understanding. It forms the deepest aspect of the workings of quantum gravity. So, Tarski's undefinability theorem reinforces the idea that quantum gravity, like any formal system that seeks to capture fundamental physical reality, cannot be self-contained and requires external structures to fully comprehend its truths \cite{Faizal20241}.
Selective Actualism is already based on such non-algorithmic, non-computational external structures, which selectively express some consequences of the theory physically. Here, we have argued that this is also true for Universal Actualism, and that the Weinberg proposal is impossible to attain due to Tarski's undefinability theorem.

G\"odel’s incompleteness theorems and Tarski’s undefinability theorem highlight the limitations of any formal, algorithmic approach to understanding quantum gravity. These theorems suggest that quantum gravity cannot be fully captured by an internal, self-contained formal system. Instead, it requires an external meta-theoretical framework to address fundamental questions about provability, truth, and consistency.  This underscores the need for an approach based on a non-algorithmic understanding to discuss quantum gravity—one that goes beyond the computational consequences of formal systems and incorporates external truth predicates for a complete description.

 \section{Chaitin's Incompleteness Theorem}
Even though formal mathematical systems are based on logic, and G\"odel's and Tarski's theorems demonstrate implicitly the limitations of  {a theory of quantum gravity purely formulated as a formal axiomatic system, 
practical calculations in quantum gravity are typically carried out using computational algorithms. To explicitly investigate these limitations, one must therefore examine the constraints within the framework of algorithmic information theory.}

Chaitin's incompleteness theorem \cite{chaitin1975theory, Chaitin2004}
is a profound result in algorithmic information theory, which results in similar fundamental limitations as G\"odel's  theorems (and is directly related to them \cite{kritchman2010surprise}). In essence, it states that in any consistent and sufficiently expressive formal system, there exists a bound on the provability of mathematical statements based on their algorithmic complexity. This theorem complements G\"odel's incompleteness theorems by demonstrating that the inability to prove certain truths arises from their informational complexity. 
It should be noted that, although algorithmic information theory demonstrates that any formal system contains infinitely many undecidable mathematical statements, only a vanishingly small fraction of such unprovable statements could plausibly relate to specific physical properties. Establishing such a correspondence requires explicit constructions, as shown for instance for specific systems  \cite{paterek2010,purcell2024}.

 We will first analyze the mathematical consequences of Chaitin's theorem for any axiomatic formalization of quantum gravity 
 \cite{Witten:1985cc, Ziaeepour:2021ubo, Faizal2024, bombelli1987spacetime, Majid:2017bul, DAriano:2016njq, Arsiwalla:2021eao}, and then discuss some physical implications. For any formal axiomatic system, Chaitin's theorem applies directly.
Accordingly, we can state that for any formal system \(
\mathcal{F}_{QG} = \mathcal{F}_{PQG} \cup \mathcal{F}_{MQG}
\), there exists a constant $K_{\mathcal{F}_{QG}}$, determined by the axioms of $\mathcal{F}_{QG}$, such that no mathematical statement $S$ with Kolmogorov complexity $K(S)$ greater than $K_{\mathcal{F}_{QG}}$ can be proven within the system. The Kolmogorov complexity $K(S)$, is defined as the length of the shortest possible program that can compute $S$ on a universal Turing machine. In other words, it is the minimal number of bits required to describe $S$ algorithmically. Thus, each mathematical statement within $\mathcal{F}_{QG}$ can be encoded as a finite binary string, allowing us to apply Chaitin's theorem to the formal structure of the theory. However, it remains unclear whether these mathematical limitations are merely a trivial consequence of the fact that
\(
\mathcal{F}_{QG} = \mathcal{F}_{PQG} \cup \mathcal{F}_{MQG}
\),
or whether they have meaningful physical implications.

It has recently been demonstrated that, due to Chaitin's theorem, the spacetime foam (i.e., Planckian quantum gravitational fluctuations of spacetime) would become undecidable \cite{Faizal2025}. It has also been argued that these fluctuations of spacetime can have detectable signatures on gravitational waves.
So, detailed descriptions of spacetime topology at the Planck scale involve complexities beyond the provability threshold of $\mathcal{F}_{QG}$, thus showing that some truths in quantum gravity, though valid, would lie outside the reach of computational derivation. This is a direct consequence of the fact that the spacetime foam is undecidable in quantum gravity as a result of Chaitin's theorem \cite{Faizal2025}.

Furthermore, any systems in quantum gravity, such as those describing the microstates of black holes, involve high complexity. The Bekenstein-Hawking entropy, for instance, scales as \( S_{BH} \sim A / (4 \ell_p^2) \), where \( A \) is the horizon area. The scaling of entropy with the black hole's surface area suggests that the number of possible microstates grows exponentially with the area, indicating a highly complex structure underlying the microstates of a black hole \cite{Balasubramanian:2024rek}. If information about individual microstates requires a Kolmogorov complexity exceeding \( K_{\mathcal{F}_{QG}} \), they will be unprovable within the formal system of quantum gravity. 
It has already been argued that quantum gravity prevents its own measurement near the Planck scale \cite{Pourhassan:2023jbs, Pourhassan2025}. In fact, these works argue that this is directly related to incompleteness theorems.

It has already been demonstrated that the Heisenberg uncertainty principle in quantum mechanics can be directly related to Chaitin's theorem \cite{Calude2005, Calude2010}. Now, it is possible to study the uncertainty principle for any quantum gravitational system, and it can directly affect the properties of spacetime. The uncertainty principle has been studied in string theory, where it has direct implications for black holes and D-branes \cite{Yoneya2000}. Similarly, the uncertainty principle has also been studied in the context of LQG \cite{Perlov2018}. 
The uncertainty principle in quantum gravity has direct implications for quantum cosmology \cite{Ohkuwa:2012cm, Ohkuwa:2012wz}.
 These results can be seen as direct consequences of Chaitin's theorem on quantum gravity.

Thus, Chaitin's theorem suggests a fundamental limitation in the axiomatic formalization of quantum gravity. It implies that there exist true mathematical statements in quantum gravity, which have a Kolmogorov complexity beyond the capacity of the formal system to prove. In other words, there are mathematical truths in quantum gravity that cannot be captured or proven by any algorithmic formal system like $\mathcal{F}_{QG}$. This aligns with the broader idea in the foundations of quantum gravity that there are aspects of the theory that may be fundamentally unprovable within any formal system, reinforcing the need for meta-theoretical frameworks to fully understand the working of quantum gravity.

 \section{Application to Physical Problems} 

We will now discuss some potential applications of such a structure beyond computation in quantum gravity. It should be noted that we will not explicitly prove the undecidability of these structures, but will instead provide physical arguments as to why they are expected to be undecidable.
We will also mention physical systems in which specific properties have been shown to correspond to mathematical undecidable statements \cite{peraleseceiza2024undecidabilityphysicsreview}. We notice however that, in most cases, these results have been obtained via the halting problem of a universal Turing machine or through other mathematical systems embedding or simulating a universal Turing machine. 
It is possible to show that the undecidability of the halting predicate and G\"odel’s incompleteness phenomena are extensionally equivalent. By encoding a G\"odel-style self-referential sentence as the source code of a program, one constructs an arithmetic statement whose provability within any consistent, computably enumerable theory would decide whether that program halts. If the theory could resolve every such statement, it would effectively compute the universal halting problem, contradicting Turing’s diagonal argument. Conversely, the inability of the theory to settle the program’s behavior mirrors its inability to prove its own consistency. Hence G\"odel’s incompleteness theorems follow almost immediately from the halting problem, affirming that no sufficiently expressive formal system can simultaneously prove its consistency and decide all halting instances \cite{Calude2021Incompleteness}.
As previously noted, undecidability arises naturally within Selective Actualism. In the following, however, we employ physical reasoning to argue that it also holds within the framework of Universal Actualism, and that computation alone is insufficient to fully account for actual physical phenomena.

We start our analysis from the black hole information paradox \cite{Almheiri2021}. 
It could be that the information stored a black hole associated with Planck scale physics, where quantum fluctuations and traditional notions of geometry appear to break down, cannot be microscopically algorithmically computed due Chaitin's incompleteness theorem, as alluded to above. In this case, the  emergence of smooth spacetime from Planck-scale physics would occur  through thermalization, where microscopic quantum gravitational degrees of freedom collectively evolve into a macroscopic geometric state. This could be beyond algorithmic computation, as in general it has been explicitly proven  that determining whether a given system thermalizes is computationally  undecidable \cite{Shiraishi2021}. 

Thermalization appears to be very important in existing quantum gravity models for understanding the emerges of classical spacetime from Planck scale physics. In string theory, via AdS/CFT, perturbations drive a rapid gravitational collapse that thermalizes the bulk into a smooth black hole horizon with well‑defined thermodynamic properties \cite{Chesler:2009cy}.  Furthermore, in fuzzball theory, thermalization emerges as the collective behavior of numerous microstate geometries that statistically reproduce the black hole’s thermal radiation spectrum \cite{Mathur:2005zp}. Similarly, in loop quantum gravity, coarse graining thermalizes discrete quantum geometries into a continuum phase that recovers classical spacetime \cite{dittrich2020coarse}.  
As it is appears to be impossible to computationally determine if given system thermalizes \cite{Shiraishi2021}, such emergence of spacetime could be uncomputable.
This might be deeply related to not only the black hole information paradox, but the emergence of spacetime itself.  So, it is possible that that a full resolution of information paradox and explanation of the emergence of spacetime can only come from a meta-theoretical framework whose principles transcend conventional computation. 

It has also been demonstrated that no computational algorithm can decide whether a quantum many-body Hamiltonian is gapped or gapless \cite{cubitt2015undecidability}. This is done by constructing a family of Hamiltonian's, whose spectral gap encodes the halting problem, so that determining the presence of a spectral gap is fundamentally undecidable.
This results holds because the halting problem \cite{turing1936computable}, which is directly related to the Chaitin's incompleteness theorem \cite{li1997introduction}, proves that no algorithm can universally decide whether an arbitrary program will halt, thus establishing a fundamental limit on computation.

This result has motivated the study of such behavior in the full renormalization group (RG) flows, which have been shown, in quantum many-body systems, to not be computable \cite{watson2022uncomputably}.  
RG flows in both string theory \cite{Callan:1985ia}, and LQG \cite{Steinhaus:2018}, play an important role in the emergence of spacetime \cite{Litim:2004, Ambjorn:2005db}. If  RG flows in quantum gravity are algorithmically uncomputable,  the emergence of spacetime from quantum gravity could be beyond algorithmic computations. It is also worth noting that certain properties of tensor networks are not computable \cite{kliesch2014matrix}. Tensor networks have been used both in string theory \cite{hayden2016holographic} and LQG \cite{Dittrich:2011zh} to explain the emergence of spacetime from quantum gravity.

Similar problems also occur in other aspects of quantum gravity. For example, it 
it has been explicitly demonstrated that as a consequence of G\"odel's incompleteness theorems, it is not  computationally possible to determine  if a given two dimensional supersymmetric  theory breaks supersymmetry \cite{tachikawa2023undecidable}. As supersymmetry is an important ingredient in superstring theories \cite{Green1984}, these results could again have important consequences in quantum gravity.  

Also, using an explicit construction of quantum spin models whose phase diagrams encode uncomputable problems, it has been shown that no general algorithm can fully determine these diagrams \cite{bausch2021uncomputability}, implying intrinsic computational limits to predicting phase transitions and understanding complex quantum matter. Now as the quantum spin model and LQG are closely related mathematically \cite{Feller:2015}, this may have important implications for LQG.  So, it is possible that the   the full phase diagram of LQG, like  quantum many-body systems, is fundamentally not computable. 

These results suggest that certain truths lie beyond the computational domain of any consistent and sufficiently expressive theory of quantum gravity. Note that the principle of sufficient reason states that every physical fact has a reason, which forms the basis of  science itself \cite{amijee2021principle, leibniz1996discourse}. The absence of an algorithmic explanation does not contradict this principle; rather, it suggests that while some explanations may be expressed in algorithmic terms, others reside in a non-algorithmic understanding, which  still provides a sufficient explanation for the phenomena in question.

In fact, several undecidable problems occur due the link between  certain properties of physical systems and the Turing's halting problem \cite{bennett1990undecidable}.   
 It has been argued using a Stewart approach that  non-algorithmic understanding can overcome such intrinsic computational limitations \cite{Stewart1991}. 
This is closely related to the Lucas-Penrose argument \cite{lucas1961minds, Penrose2011, Penrose1990-PENTNM, hameroff2014consciousness, lucas_penrose_2023}), indicating that there is something intrinsically non-computational in nature, and its description in terms of computational algorithms is rather limited. 

The Novikov self-consistency principle \cite{Friedman1990} provides another compelling example of non-algorithmic understanding.  {Originally formulated in the context of time travel and closed timelike curves in general relativity, it posits that events within regions of spacetime containing causal loops must be self-consistent. So, any action taken by a time traveler must have already been part of history, ensuring that the timeline remains consistent and free from contradictions \cite{novikov1989}. This principle cannot be computationally derived from general relativity axioms but is an intuitive addition that resolves paradoxes due to self-referentiality. Similarly, in quantum gravity, principles inspired by the Novikov self-consistency framework may enable the resolution of truths that exceed the provability limits of formal systems due to self-referentiality, offering a broader understanding of phenomena that cannot be computational derived.

Gravitationally induced objective collapse, which addresses the measurement problem in quantum mechanics using an objective physical process tied to gravitational effects, also exemplifies the application of non-algorithmic understanding  \cite{Penrose1996, Diosi1987}.  In this framework, measurement occurs independently of an observer, and is neither random nor computational.} In fact,  it has been explicitly demonstrated that even  quantum logic is inherently undecidable due to incompleteness theorems  \cite{Fritz2021, Eisert2012, vandenNest2008measurement, lloyd1993quantum}. So,  an approach based on non-algorithmic understanding \cite{Penrose1996, Diosi1987} produces a better physical explanation of the measurement problem.
It is well established that there are limitations to computations in various problems of physics \cite{peraleseceiza2024undecidabilityphysicsreview}. Here we have argued that these similarly imply it impossible for quantum gravity based on computations to be both consistent and complete.  Here, it is important to note that even if some specific phenomena discussed here are not undecidable, the general arguments presented imply that there will definitely be some undecidable phenomena. Hence, this problem will inevitably arise, and we need to go beyond a computational framework to address it.

To address the inherent limitations of formal systems in quantum gravity, a meta-theory of everything  \( \mathcal{M}_{ToE} \) is needed, which contains non-algorithmic understanding. This meta-theory of everything would extend the computational-based formal system \( \mathcal{F}_{QG} \) by introducing a truth predicate \( T(x) \), enabling the recognition of truths beyond formal provability. Specifically, \( \mathcal{M}_{ToE} \) would allow for the acceptance of G\"odel statements \( G \) as true, the resolution of undefinable truths, and the acknowledgment of high-complexity mathematical statements \( S \) with \( K(S) > K_{\mathcal{F}_{QG}} \). This meta-theory would satisfy:
$
T(G) = 1 \quad \text{for } G \text{ true but unprovable in } \mathcal{F}_{QG}.
$
 {By transcending the algorithmic limits of \( \mathcal{F}_{QG} \), \( \mathcal{M}_{ToE} \) would provide a framework to explore truths that lie beyond computation. Inspired by principles such as Novikov self-consistency, and similar principles based on non-algorithmic understanding, this approach could integrate rigorous computations with non-algorithmic insights. It could potentially resolve challenges like the black hole information paradox and the complexities of Planck-scale physics.  }
 
 Considerations of quantum gravity have suggested that physics should evolve from a focus on `it' (i.e. matter fields moving in spacetime) to a focus on `bit', i.e., information being more fundamental.  The considerations we present here suggest that neither `it', nor `bit' is fundamental. Namely, a non-algorithmic understanding should operate at a fundamental level in any theory of everything.    

\newpage
\section*{Acknowledgment} We would like to thank İzzet Sakallı, Salman Sajad Wani, and Aatif Kaisar Khan
 for useful discussions. We would also like to thank Aatif Kaisar Khan for sharing with us important paper on undecidability.    We want to thank Stephen Hawking for his discussion on G\"odel’s theorems  and the end of physics, which motivated the current works.  We would also like to thank Roger Penrose for his explanation of G\"odel’s theorems and Lucas-Penrose argument, which forms the basis of meta-theoretical perspective based on non-algorithmic understanding developed here. 
\bibliographystyle{utphys}
\bibliography{references}

\providecommand{\href}[2]{#2}\begingroup\raggedright\begin{thebibliography}{10}

\bibitem{Einstein1915}
A.~Einstein, ``{Die Feldgleichungen der Gravitation},''
  \href{http://dx.doi.org/https://doi.org/10.1007/978-3-322-83770-7_10}{{\em
  Sitzungsberichte der Preussischen Akademie der Wissenschaften zu Berlin}
  (1915)  }.

\bibitem{Abbott2016}
B.~P. Abbott {\em et al.}, ``{Observation of Gravitational Waves from a Binary
  Black Hole Merger},''
\href{http://dx.doi.org/10.1103/PhysRevLett.116.061102}{{\em Physical Review
  Letters} {\bf 116} (2016)  061102}.

\bibitem{Penrose1965}
R.~Penrose, ``{Gravitational Collapse and Space-Time Singularities},''
\href{http://dx.doi.org/10.1103/PhysRevLett.14.57}{{\em Physical Review
  Letters} {\bf 14} (1965)  57}.

\bibitem{Hawking1970}
S.~Hawking and R.~Penrose, ``{The Singularities of Gravitational Collapse and
  Cosmology},'' \href{http://dx.doi.org/10.1098/rspa.1970.0021}{{\em
  Proceedings of the Royal Society A} {\bf 314} (1970)  529}.

\bibitem{Arnold92}
V.~I. Arnold, {\em Catastrophe Theory}.
\newblock Springer Berlin, Heidelberg, 1992.
\newblock \url{https://link.springer.com/book/10.1007/978-3-642-58124-3}.

\bibitem{Berry2023}
M.~Berry, ``The singularities of light: intensity, phase, polarisation,''
  \href{http://dx.doi.org/https://doi.org/10.1038/s41377-023-01270-8}{{\em
  Light Sci. Appl.} {\bf 12} (2023)  238}.

\bibitem{Faccio2016}
F.~Marino, C.~Maitland, D.~Vocke, O.~Ortolan, and D.~Faccio, ``{Emergent
  geometries and nonlinear-wave dynamics in photon fluids},''
  \href{http://dx.doi.org/https://doi.org/10.1038/srep23282}{{\em Scientific
  Reports} {\bf 6} (2016)  23282}.

\bibitem{Braunstein:2023jpo}
S.~L. Braunstein, M.~Faizal, L.~M. Krauss, F.~Marino, and N.~A. Shah,
  ``{Analogue simulations of quantum gravity with fluids},''
  \href{http://dx.doi.org/10.1038/s42254-023-00630-y}{{\em Nature Rev. Phys.}
  {\bf 5} (2023) no.~10, 612--622}, \href{http://arxiv.org/abs/2402.16136}{{\tt
  arXiv:2402.16136 [gr-qc]}}.

\bibitem{Bojowald2001}
M.~Bojowald, ``{Absence of Singularity in Loop Quantum Cosmology},''
  \href{http://dx.doi.org/10.1103/PhysRevLett.86.5227}{{\em Physical Review
  Letters} {\bf 86} (2001)  5227}.

\bibitem{Ashtekar2006}
A.~Ashtekar, T.~Pawlowski, and P.~Singh, ``{Quantum Nature of the Big Bang:
  Improved Dynamics},''
  \href{http://dx.doi.org/10.1103/PhysRevD.74.084003}{{\em Phys. Rev. D} {\bf
  74} (2006)  084003}.

\bibitem{Mathur2005}
S.~D. Mathur, ``{The Fuzzball Proposal for Black Holes: An Elementary
  Review},''
\href{http://dx.doi.org/10.1002/prop.200410203}{{\em Fortschritte der Physik}
  {\bf 53} (2005)  793}.

\bibitem{Mathur2008}
S.~D. Mathur, ``{Tunneling into fuzzball states},''
  \href{http://dx.doi.org/10.1007/s10714-009-0837-3}{{\em Gen. Rel. Grav.} {\bf
  42} (2010)  113--118}, \href{http://arxiv.org/abs/0805.3716}{{\tt
  arXiv:0805.3716 [hep-th]}}.

\bibitem{Hohm2010}
O.~Hohm, C.~Hull, and B.~Zwiebach, ``{Generalized Metric Formulation of Double
  Field Theory},'' \href{http://dx.doi.org/10.1007/JHEP08(2010)008}{{\em
  Journal of High Energy Physics} {\bf 08} (2010)  008},
\href{http://arxiv.org/abs/1006.4823}{{\tt arXiv:1006.4823 [hep-th]}}.

\bibitem{Hull2005}
C.~M. Hull, ``{A Geometry for Non-Geometric String Backgrounds},''
  \href{http://dx.doi.org/10.1088/1126-6708/2005/10/065}{{\em Journal of High
  Energy Physics} {\bf 10} (2005)  065},
\href{http://arxiv.org/abs/hep-th/0406102}{{\tt arXiv:hep-th/0406102
  [hep-th]}}.

\bibitem{Perez2013}
A.~Perez, ``{The Spin Foam Approach to Quantum Gravity},''
\href{http://dx.doi.org/10.12942/lrr-2013-3}{{\em Living Reviews in Relativity}
  {\bf 16} (2013)  3}.

\bibitem{Jafferis:2022crx}
D.~Jafferis, A.~Zlokapa, J.~D. Lykken, D.~K. Kolchmeyer, S.~I. Davis, N.~Lauk,
  H.~Neven, and M.~Spiropulu, ``{Traversable wormhole dynamics on a quantum
  processor},'' \href{http://dx.doi.org/10.1038/s41586-022-05424-3}{{\em
  Nature} {\bf 612} (2022) no.~7938, 51--55}.

\bibitem{VanRaamsdonk:2020ydg}
M.~Van~Raamsdonk, ``{Spacetime from bits},''
  \href{http://dx.doi.org/10.1126/science.aay9560}{{\em Science} {\bf 370}
  (2020) no.~6513, 198--202}.

\bibitem{Makela:2019vgf}
J.~M\"akel\"a, ``{Wheeler\textquoteright{}s it from bit proposal in loop
  quantum gravity},'' \href{http://dx.doi.org/10.1142/S0218271819501293}{{\em
  Int. J. Mod. Phys. D} {\bf 28} (2019) no.~10, 1950129},
  \href{http://arxiv.org/abs/1905.02611}{{\tt arXiv:1905.02611 [gr-qc]}}.

\bibitem{Wheeler1990}
J.~A. Wheeler, ``Information, physics, quantum: The search for links,'' {\em
  Complexity, Entropy, and the Physics of Information} (1990)  .

\bibitem{weinberg2011dreams}
S.~Weinberg, {\em Dreams of a Final Theory: The Scientist's Search for the
  Ultimate Laws of Nature}.
\newblock Vintage books. Knopf Doubleday Publishing Group, 1994.
\newblock \url{https://books.google.co.in/books?id=Qd0MEtsBr7oC}.

\bibitem{tegmark2008mu}
M.~Tegmark, ``The mathematical universe,''
  \href{http://dx.doi.org/10.1007/s10701-007-9186-9}{{\em Foundations of
  Physics} {\bf 38} (2008) no.~2, 101--150},
  \href{http://arxiv.org/abs/0704.0646}{{\tt arXiv:0704.0646 [gr-qc]}}.
  \url{https://doi.org/10.1007/s10701-007-9186-9}.

\bibitem{schmidhuber1997computer}
J.~Schmidhuber, \href{http://dx.doi.org/10.1007/BFb0052088}{``A computer
  scientist’s view of life, the universe, and everything,''} in {\em
  Foundations of Computer Science: Potential -- Theory -- Cognition}, vol.~1337
  of {\em Lecture Notes in Computer Science}, pp.~201--208.
\newblock Springer, Berlin, Heidelberg, 1997.

\bibitem{Green1987}
M.~B. Green, J.~H. Schwarz, and E.~Witten,
  \href{http://dx.doi.org/https://doi.org/10.1017/CBO9781139248563}{{\em
  Superstring Theory}}.
\newblock Cambridge University Press, 1987.

\bibitem{Polchinski1998}
J.~Polchinski,
  \href{http://dx.doi.org/https://doi.org/10.1017/CBO9780511816079}{{\em String
  Theory}}.
\newblock Cambridge University Press, 1998.

\bibitem{Rovelli2004}
C.~Rovelli,
  \href{http://dx.doi.org/https://doi.org/10.1017/CBO9780511755804}{{\em
  {Quantum Gravity}}}.
\newblock Cambridge University Press, Cambridge, UK, 2004.

\bibitem{Thiemann2007}
T.~Thiemann,
  \href{http://dx.doi.org/https://doi.org/10.1017/CBO9780511755682}{{\em Modern
  Canonical Quantum General Relativity}}.
\newblock Cambridge University Press, 2007.

\bibitem{Witten:1985cc}
E.~Witten, ``{Noncommutative Geometry and String Field Theory},''
  \href{http://dx.doi.org/10.1016/0550-3213(86)90155-0}{{\em Nucl. Phys. B}
  {\bf 268} (1986)  253--294}.

\bibitem{Ziaeepour:2021ubo}
H.~Ziaeepour, ``{Comparing Quantum Gravity Models: String Theory, Loop Quantum
  Gravity, and Entanglement Gravity versus SU(\ensuremath{\infty})-QGR},''
  \href{http://dx.doi.org/10.3390/sym14010058}{{\em Symmetry} {\bf 14} (2022)
  58}, \href{http://arxiv.org/abs/2109.05757}{{\tt arXiv:2109.05757 [gr-qc]}}.

\bibitem{Faizal2024}
M.~Faizal, A.~Shabir, and A.~K. Khan, ``{Consequences of G\"odel theorems on
  third quantized theories like string field theory and group field theory},''
  \href{http://dx.doi.org/10.1016/j.nuclphysb.2024.116774}{{\em Nucl. Phys. B}
  {\bf 1010} (2025)  116774}, \href{http://arxiv.org/abs/2407.12313}{{\tt
  arXiv:2407.12313 [hep-th]}}.

\bibitem{bombelli1987spacetime}
L.~Bombelli, J.~Lee, D.~Meyer, and R.~D. Sorkin, ``Spacetime as a causal set,''
  \href{http://dx.doi.org/10.1103/PhysRevLett.59.521}{{\em Physical Review
  Letters} {\bf 59} (1987) no.~5, 521--524}.

\bibitem{Majid:2017bul}
S.~Majid, ``{On the emergence of the structure of Physics},''
  \href{http://dx.doi.org/10.1098/rsta.2017.0231}{{\em Phil. Trans. Roy. Soc.
  Lond. A} {\bf 376} (2018)  0231}, \href{http://arxiv.org/abs/1711.00556}{{\tt
  arXiv:1711.00556 [math-ph]}}.

\bibitem{DAriano:2016njq}
G.~M. D'Ariano, ``{Physics Without Physics: The Power of
  Information-theoretical Principles},''
  \href{http://dx.doi.org/10.1007/s10773-016-3172-y}{{\em Int. J. Theor. Phys.}
  {\bf 56} (2017) no.~1, 97--128}, \href{http://arxiv.org/abs/1701.06309}{{\tt
  arXiv:1701.06309 [quant-ph]}}.

\bibitem{Arsiwalla:2021eao}
X.~D. Arsiwalla and J.~Gorard, ``{Pregeometric Spaces from Wolfram Model
  Rewriting Systems as Homotopy Types},''
  \href{http://dx.doi.org/10.1007/s10773-024-05576-0}{{\em Int. J. Theor.
  Phys.} {\bf 63} (2024) no.~4, 83},
  \href{http://arxiv.org/abs/2111.03460}{{\tt arXiv:2111.03460 [math.CT]}}.

\bibitem{Faizal2023}
M.~Faizal, ``{The end of space\textendash{}time},''
  \href{http://dx.doi.org/10.1142/S0217751X23501889}{{\em Int. J. Mod. Phys. A}
  {\bf 38} (2023) no.~35n36, 2350188},
  \href{http://arxiv.org/abs/2403.10694}{{\tt arXiv:2403.10694 [gr-qc]}}.

\bibitem{Seiberg2006}
N.~Seiberg, \href{http://dx.doi.org/10.1142/9789812706768_0005}{``{Emergent
  spacetime},''} in {\em {23rd Solvay Conference in Physics: The Quantum
  Structure of Space and Time}}, pp.~163--178.
\newblock 1, 2006.
\newblock \href{http://arxiv.org/abs/hep-th/0601234}{{\tt
  arXiv:hep-th/0601234}}.

\bibitem{Green1984}
M.~B. Green and J.~H. Schwarz, ``Anomaly cancellation in supersymmetric d=10
  gauge theory,'' \href{http://dx.doi.org/10.1016/0370-2693(84)91565-X}{{\em
  Physics Letters B} {\bf 149} (1984)  117}.

\bibitem{Ashtekar1986}
A.~Ashtekar, ``New variables for classical and quantum gravity,''
  \href{http://dx.doi.org/https://doi.org/10.1103/PhysRevLett.57.2244}{{\em
  Physical Review Letters} {\bf 57} (1986)  2244}.

\bibitem{Khoury:2001wf}
J.~Khoury, B.~A. Ovrut, P.~J. Steinhardt, and N.~Turok, ``{The Ekpyrotic
  universe: Colliding branes and the origin of the hot big bang},''
  \href{http://dx.doi.org/10.1103/PhysRevD.64.123522}{{\em Phys. Rev. D} {\bf
  64} (2001)  123522}, \href{http://arxiv.org/abs/hep-th/0103239}{{\tt
  arXiv:hep-th/0103239}}.

\bibitem{Gaona-Reyes:2024qcc}
J.~L. Gaona-Reyes, L.~Men\'endez-Pidal, M.~Faizal, and M.~Carlesso,
  ``{Spontaneous collapse models lead to the emergence of classicality of the
  Universe},'' \href{http://dx.doi.org/10.1007/JHEP02(2024)193}{{\em JHEP} {\bf
  02} (2024)  193}, \href{http://arxiv.org/abs/2401.08269}{{\tt
  arXiv:2401.08269 [gr-qc]}}.

\bibitem{Penrose1996}
R.~Penrose, ``On gravity’s role in quantum state reduction,''
  \href{http://dx.doi.org/10.1007/BF02105068}{{\em General Relativity and
  Gravitation} {\bf 28} (1996)  581}.

\bibitem{Diosi1987}
L.~Diósi, ``A universal master equation for the gravitational violation of
  quantum mechanics,''
  \href{http://dx.doi.org/https://doi.org/10.1016/0375-9601(87)90681-5}{{\em
  Physics Letters A} {\bf 120} (1987)  377}.

\bibitem{novikov1989}
I.~D. Novikov, ``Time machine and self-consistent evolution in problems with
  self-interaction,''
  \href{http://dx.doi.org/https://doi.org/10.1103/PhysRevD.45.1989}{{\em Phys.
  Rev. D} {\bf 45} (1992) no.~11, }.

\bibitem{Pourhassan:2023jbs}
B.~Pourhassan, X.~Shi, S.~S. Wani, Saif-Al-Khawari, F.~Kazemian, I.~Sakall\i{},
  N.~A. Shah, and M.~Faizal, ``{Information Theoretical Approach to Detecting
  Quantum Gravitational Corrections},''
  \href{http://dx.doi.org/10.1007/JHEP02(2025)109}{{\em JHEP} {\bf 25} (2025)
  109}, \href{http://arxiv.org/abs/2310.12878}{{\tt arXiv:2310.12878
  [hep-th]}}.

\bibitem{Pourhassan2025}
B.~Pourhassan, X.~Shi, S.~S. Wani, S.~Al-Kuwari, {\.I}.~Sakall{\i}, N.~A. Shah,
  M.~Faizal, and A.~Shabir,
  \href{http://dx.doi.org/10.1016/j.aop.2024.169983}{``Quantum gravitational
  corrections to a {Kerr} black hole using {Topos} theory,''{\em Annals of
  Physics} {\bf 477} (June, 2025)  169983}.

\bibitem{lucas1961minds}
J.~R. Lucas, ``Minds, machines and gödel,''
  \href{http://dx.doi.org/10.1017/S0031819100057983}{{\em Philosophy} {\bf 36}
  (1961) no.~137, 112--127}.

\bibitem{Penrose2011}
R.~Penrose, ``G\"{o}del, the mind, and the laws of physics,'' in {\em Kurt
  G\"{o}del's and the foundations of mathematics: horizons of truth}, p.~339.
\newblock Cambridge University Press, 2011.

\bibitem{Penrose1990-PENTNM}
R.~Penrose, ``The nonalgorithmic mind,''
  \href{http://dx.doi.org/10.1017/s0140525x0008105x}{{\em Behavioral and Brain
  Sciences} {\bf 13} (1990) no.~4, 692--}.

\bibitem{hameroff2014consciousness}
S.~Hameroff and R.~Penrose, ``Consciousness in the universe: A review of the
  'orch or' theory,'' \href{http://dx.doi.org/10.1016/j.plrev.2013.08.002}{{\em
  Physics of Life Reviews} {\bf 11} (2014) no.~1, 39--78}.
  \url{https://doi.org/10.1016/j.plrev.2013.08.002}.

\bibitem{lucas_penrose_2023}
J.~P. S., \href{http://dx.doi.org/10.1007/978-3-031-64217-3_7}{``The
  lucas–penrose arguments,''} in {\em The Argument of Mathematics},
  p.~Chapter 7.
\newblock Springer, 2023.
\newblock \url{https://link.springer.com/chapter/10.1007/978-3-031-64217-3_7}.

\bibitem{Godel1931}
K.~G\"{o}del, ``\"{U}ber formal unentscheidbare s\"{a}tze der principia
  mathematica und verwandter systeme i,''
  \href{http://dx.doi.org/https://doi.org/10.1007/BF01700692}{{\em Monatshefte
  f\"{u}r Mathematik} {\bf 38} (1931) no.~1, 173--198}.

\bibitem{Smith2007}
P.~Smith,
  \href{http://dx.doi.org/https://doi.org/10.1017/CBO9781139149105}{{\em {An
  Introduction to Gödel's Theorems}}}.
\newblock Cambridge University Press, Cambridge, 2nd~ed., 2007.

\bibitem{Faizal20240}
M.~Faizal, A.~Shabir, and A.~K. Khan, ``{Consequences of
  G\"odel\textquoteright{}s Theorems on Quantum Gravity},''
  \href{http://dx.doi.org/10.1007/s10773-024-05818-1}{{\em Int. J. Theor.
  Phys.} {\bf 63} (2024) no.~11, 290}.

\bibitem{Fritz2021}
T.~Fritz, ``Quantum logic is undecidable,'' {\em Archive for Mathematical
  Logic} {\bf 60} (2021)  329--341.

\bibitem{Eisert2012}
J.~Eisert, M.~P. Müller, and C.~Gogolin, ``Quantum measurement occurrence is
  undecidable,'' \href{http://dx.doi.org/10.1103/PhysRevLett.108.260501}{{\em
  Physical Review Letters} {\bf 108} (2012)  260501}.

\bibitem{vandenNest2008measurement}
M.~Van~den Nest and H.~J. Briegel, ``Measurement-based quantum computation and
  undecidable logic,'' \href{http://dx.doi.org/10.1007/s10701-008-9212-6}{{\em
  Foundations of Physics} {\bf 38} (2008)  448--457}.
  \url{https://link.springer.com/article/10.1007/s10701-008-9212-6}.

\bibitem{lloyd1993quantum}
S.~Lloyd, ``Quantum-mechanical computers and uncomputability,''
  \href{http://dx.doi.org/10.1103/PhysRevLett.71.943}{{\em Physical Review
  Letters} {\bf 71} (1993) no.~6, 943--946}.
  \url{https://link.aps.org/doi/10.1103/PhysRevLett.71.943}.

\bibitem{Panagiotopoulos2023}
A.~Panagiotopoulos, G.~Sparling, and M.~Christodoulou, ``Incompleteness
  theorems for observables in general relativity,''
  \href{http://dx.doi.org/10.1103/PhysRevLett.131.171402}{{\em Physical Review
  Letters} {\bf 131} (2023) no.~17, 171402}.

\bibitem{Tarski1933}
A.~Tarski, ``{Pojecie Prawdy w Jezykach Nauk Dedukcyjnych (The Concept of Truth
  in the Languages of the Deductive Sciences)},'' {\em Prace Towarzystwa
  Naukowego Warszawskiego, Wydział III} {\bf 34} (1933)  .

\bibitem{Tarski1983}
A.~Tarski, \href{http://dx.doi.org/http://dx.doi.org/10.2307/2275031}{{\em
  {Logic, Semantics, Metamathematics: Papers from 1923 to 1938}}}.
\newblock Hackett Publishing Company, Indianapolis, 1983.

\bibitem{Faizal20241}
M.~Faizal, A.~Shabir, and A.~K. Khan, ``{Implications of Tarski's
  undefinability theorem on the Theory of Everything},''
  \href{http://dx.doi.org/10.1209/0295-5075/ad80c2}{{\em EPL} {\bf 148} (2024)
  no.~3, 39001}, \href{http://arxiv.org/abs/2410.10903}{{\tt arXiv:2410.10903
  [physics.hist-ph]}}.

\bibitem{chaitin1975theory}
G.~J. Chaitin, ``A theory of program size formally identical to information
  theory,'' \href{http://dx.doi.org/10.1145/321892.321894}{{\em Journal of the
  ACM} {\bf 22} (1975) no.~3, 329--340}.
  \url{https://doi.org/10.1145/321892.321894}.

\bibitem{Chaitin2004}
G.~J. Chaitin,
  \href{http://dx.doi.org/https://doi.org/10.48550/arXiv.math/0404335}{{\em
  {Meta Math!: The Quest for Omega}}}.
\newblock Pantheon Books, New York, 2004.

\bibitem{kritchman2010surprise}
S.~Kritchman and R.~Raz, ``The surprise examination paradox and the second
  incompleteness theorem,''
  \href{http://dx.doi.org/10.48550/arXiv.1011.4974}{{\em Notices of the AMS}
  {\bf 57} (2010) no.~11, 1454}.

\bibitem{paterek2010}
T.~Paterek, J.~Kofler, R.~Prevedel, P.~Klimek, M.~Aspelmeyer, A.~Zeilinger, and
  C.~Brukner, ``{Logical independence and quantum randomness},''
  \href{http://dx.doi.org/https://doi.org/10.1088/1367-2630/12/1/013019}{{\em
  New J. Phys. 12, 013019 (2010)} {\bf 12} (2010)  013019}.

\bibitem{purcell2024}
J.~Purcell, Z.~Li, and T.~Cubitt, ``{Chaitin Phase Transitions},''
  \href{http://arxiv.org/abs/2410.02600}{{\tt arXiv:2410.02600 [quant-ph]}}.

\bibitem{Faizal2025}
M.~Faizal, A.~Shabir, and A.~K. Khan, ``Application of chaitin’s
  incompleteness theorem to quantum gravity,''
  \href{http://dx.doi.org/10.1007/s10773-025-05545-2}{{\em International
  Journal of Theoretical Physics} {\bf 64} (2025) no.~7, 194}.

\bibitem{Balasubramanian:2024rek}
V.~Balasubramanian, B.~Craps, J.~Hernandez, M.~Khramtsov, and M.~Knysh,
  ``{Counting microstates of out-of-equilibrium black hole fluctuations},''
  \href{http://dx.doi.org/10.1007/JHEP06(2025)083}{{\em JHEP} {\bf 06} (2025)
  083}.

\bibitem{Calude2005}
C.~Calude and M.~Stay, ``From heisenberg to gödel via chaitin,''
  \href{http://dx.doi.org/10.1007/s10773-005-7081-8}{{\em International Journal
  of Theoretical Physics} {\bf 44} (2005) no.~7, 1053--1065}.
  \url{https://doi.org/10.48550/arXiv.quant-ph/0402197}.

\bibitem{Calude2010}
C.~S. Calude, M.~J. Dinneen, M.~Dumitrescu, and K.~Svozil, ``Experimental
  evidence of quantum randomness incomputability,''
  \href{http://dx.doi.org/10.1103/PhysRevA.82.022102}{{\em Phys. Rev. A} {\bf
  82} (2010)  022102}.

\bibitem{Yoneya2000}
T.~Yoneya, ``String theory and the space-time uncertainty principle,''
  \href{http://dx.doi.org/10.1143/PTP.103.1081}{{\em Progress of Theoretical
  Physics} {\bf 103} (2000) no.~5, 1081--1125}.

\bibitem{Perlov2018}
L.~Perlov, ``Uncertainty principle in loop quantum cosmology by moyal
  formalism,'' \href{http://dx.doi.org/10.1063/1.5013206}{{\em Journal of
  Mathematical Physics} {\bf 59} (2018) no.~3, 032304}.

\bibitem{Ohkuwa:2012cm}
Y.~Ohkuwa and Y.~Ezawa, ``Third quantization of $f(r)$-type gravity ii -
  general $f(r)$ case,''
  \href{http://dx.doi.org/10.1088/0264-9381/30/23/235015}{{\em Class. Quant.
  Grav.} {\bf 30} (2013)  235015}.

\bibitem{Ohkuwa:2012wz}
Y.~Ohkuwa and Y.~Ezawa, ``Third quantization of $f(r)$-type gravity,''
  \href{http://dx.doi.org/10.1088/0264-9381/29/21/215004}{{\em Class. Quant.
  Grav.} {\bf 29} (2012)  215004}.

\bibitem{peraleseceiza2024undecidabilityphysicsreview}
Álvaro Perales-Eceiza, T.~Cubitt, M.~Gu, D.~Pérez-García, and M.~M. Wolf,
  ``Undecidability in physics: a review,'' 2024.
\newblock \url{https://arxiv.org/abs/2410.16532}.

\bibitem{Calude2021Incompleteness}
C.~S. Calude, ``Incompleteness and the halting problem,''
  \href{http://dx.doi.org/10.1007/s11225-021-09945-2}{{\em Studia Logica} {\bf
  109} (2021) no.~5, 1053--1067}.

\bibitem{Almheiri2021}
A.~Almheiri, T.~Hartman, J.~Maldacena, E.~Shaghoulian, and A.~Tajdini, ``The
  entropy of hawking radiation,''
  \href{http://dx.doi.org/10.1103/RevModPhys.93.035002}{{\em Reviews of Modern
  Physics} {\bf 93} (2021) no.~3, 035002}.

\bibitem{Shiraishi2021}
N.~Shiraishi and K.~Matsumoto, ``Undecidability in quantum thermalization,''
  \href{http://dx.doi.org/10.1038/s41467-021-25053-0}{{\em Nature
  Communications} {\bf 12} (2021)  5084}.

\bibitem{Chesler:2009cy}
P.~M. Chesler and L.~G. Yaffe, ``Horizon formation and far-from-equilibrium
  isotropization in a supersymmetric yang-mills plasma,''
  \href{http://dx.doi.org/10.1103/PhysRevLett.102.211601}{{\em Phys. Rev.
  Lett.} {\bf 102} (2009)  211601}, \href{http://arxiv.org/abs/0906.4426}{{\tt
  arXiv:0906.4426 [hep-th]}}.

\bibitem{Mathur:2005zp}
S.~D. Mathur, ``The fuzzball proposal for black holes: An elementary review,''
  \href{http://dx.doi.org/10.1002/prop.200410203}{{\em Fortsch. Phys.} {\bf 53}
  (2005)  793--827}, \href{http://arxiv.org/abs/hep-th/0502050}{{\tt
  arXiv:hep-th/0502050 [hep-th]}}.

\bibitem{dittrich2020coarse}
S.~Steinhaus, ``Coarse graining spin foam quantum gravity—a review,''
  \href{http://dx.doi.org/10.3389/fphy.2020.00295}{{\em Frontiers in Physics}
  {\bf 8} (2020)  }.

\bibitem{cubitt2015undecidability}
T.~S. Cubitt, D.~Perez-Garcia, and M.~M. Wolf, ``Undecidability of the spectral
  gap,'' \href{http://dx.doi.org/10.1038/nature16071}{{\em Nature} {\bf 528}
  (2015) no.~7581, 207--211}.

\bibitem{turing1936computable}
A.~M. Turing, ``On computable numbers, with an application to the
  entscheidungsproblem,''
  \href{http://dx.doi.org/10.1112/plms/s2-42.1.230}{{\em Proceedings of the
  London Mathematical Society} {\bf s2-42} (1937) no.~1, 230}.

\bibitem{li1997introduction}
M.~Li and P.~Vitányi, \href{http://dx.doi.org/10.1007/978-3-030-11298-1}{{\em
  An Introduction to Kolmogorov Complexity and Its Applications}}.
\newblock Springer, 2019.

\bibitem{watson2022uncomputably}
J.~D. Watson, E.~Onorati, and T.~S. Cubitt, ``Uncomputably complex
  renormalisation group flows,''
  \href{http://dx.doi.org/10.1038/s41467-022-35179-4}{{\em Nature
  Communications} {\bf 13} (2022) no.~1, 7618}.
  \url{https://doi.org/10.1038/s41467-022-35179-4}.

\bibitem{Callan:1985ia}
C.~G. Callan, D.~Friedan, E.~J. Martinec, and M.~J. Perry, ``Strings in
  background fields,''
  \href{http://dx.doi.org/10.1016/0550-3213(85)90506-1}{{\em Nucl. Phys. B}
  {\bf 262} (1985)  593--609}.

\bibitem{Steinhaus:2018}
S.~Steinhaus and J.~Thürigen, ``Emergence of spacetime in a restricted
  spin-foam model,'' \href{http://dx.doi.org/10.1103/PhysRevD.98.026013}{{\em
  Phys. Rev. D} {\bf 98} (2018) no.~2, 026013},
  \href{http://arxiv.org/abs/1803.10289}{{\tt arXiv:1803.10289 [gr-qc]}}.

\bibitem{Litim:2004}
D.~Litim, ``Fixed points of quantum gravity,''
  \href{http://dx.doi.org/10.1103/PhysRevLett.92.201301}{{\em Phys. Rev. Lett.}
  {\bf 92} (2004)  201301}.

\bibitem{Ambjorn:2005db}
J.~Ambj\o{}rn, J.~Jurkiewicz, and R.~Loll,
  \href{http://dx.doi.org/10.1103/PhysRevLett.95.171301}{``The spectral
  dimension of the universe is scale dependent,''{\em Phys. Rev. Lett.} {\bf
  95} (Oct, 2005)  171301}.

\bibitem{kliesch2014matrix}
M.~Kliesch, D.~Gross, and J.~Eisert, ``Matrix-product operators and states:
  Np-hardness and undecidability,''
  \href{http://dx.doi.org/10.1103/PhysRevLett.113.160503}{{\em Physical Review
  Letters} {\bf 113} (2014) no.~16, 160503}.
  \url{https://link.aps.org/doi/10.1103/PhysRevLett.113.160503}.

\bibitem{hayden2016holographic}
P.~Hayden, S.~Nezami, X.-L. Qi, N.~Thomas, M.~Walter, and Z.~Yang,
  ``Holographic duality from random tensor networks,''
  \href{http://dx.doi.org/10.1007/JHEP11(2016)009}{{\em Journal of High Energy
  Physics} {\bf 2016} (2016)  009}, \href{http://arxiv.org/abs/1601.01694}{{\tt
  arXiv:1601.01694 [hep-th]}}.

\bibitem{Dittrich:2011zh}
B.~Dittrich, F.~C. Eckert, and M.~Martin-Benito, ``Coarse graining methods for
  spin net and spin foam models,''
  \href{http://dx.doi.org/10.1088/1367-2630/14/3/035008}{{\em New J. Phys.}
  {\bf 14} (2012)  035008}, \href{http://arxiv.org/abs/1109.4927}{{\tt
  arXiv:1109.4927 [gr-qc]}}.

\bibitem{tachikawa2023undecidable}
Y.~Tachikawa, ``Undecidable problems in quantum field theory,''
  \href{http://dx.doi.org/10.1007/s10773-023-05357-1}{{\em International
  Journal of Theoretical Physics} {\bf 62} (2023)  199}.
  \url{https://doi.org/10.1007/s10773-023-05357-1}.

\bibitem{bausch2021uncomputability}
J.~Bausch, T.~S. Cubitt, and J.~D. Watson, ``Uncomputability of phase
  diagrams,'' \href{http://dx.doi.org/10.1038/s41467-020-20504-6}{{\em Nature
  Communications} {\bf 12} (2021)  452}.
  \url{https://doi.org/10.1038/s41467-020-20504-6}.

\bibitem{Feller:2015}
A.~Feller and E.~R. Livine, ``Ising spin network states for loop quantum
  gravity: A toy model for phase transitions,''
  \href{http://dx.doi.org/10.1088/0264-9381/33/6/065005}{{\em Class. Quant.
  Grav.} {\bf 33} (2016) no.~6, 065005},
  \href{http://arxiv.org/abs/1509.05297}{{\tt arXiv:1509.05297 [gr-qc]}}.

\bibitem{amijee2021principle}
F.~Amijee, \href{http://dx.doi.org/10.1007/978-3-319-20791-9_593-1}{``Principle
  of sufficient reason,''} in {\em Encyclopedia of Early Modern Philosophy and
  the Sciences}, D.~Jalobeanu and C.~T. Wolfe, eds.
\newblock Springer, 2021.

\bibitem{leibniz1996discourse}
G.~W. Leibniz, {\em Discourse on Metaphysics}.
\newblock Hackett Publishing Company, Indianapolis, 1996.
\newblock A seminal work where Leibniz famously asserts that "nothing happens
  without a reason.".

\bibitem{bennett1990undecidable}
C.~H. Bennett, ``Undecidable dynamics,''
  \href{http://dx.doi.org/10.1038/346606a0}{{\em Nature} {\bf 346} (1990)
  606--607}. \url{https://doi.org/10.1038/346606a0}.

\bibitem{Stewart1991}
I.~Stewart, ``Deciding the undecidable,''
  \href{http://dx.doi.org/10.1038/352664a0}{{\em Nature} {\bf 352} (1991)
  664--665}.

\bibitem{Friedman1990}
J.~L. Friedman, M.~S. Morris, I.~D. Novikov, F.~Echeverria, G.~Klinkhammer,
  K.~S. Thorne, and U.~Yurtsever, ``{Cauchy problem in spacetimes with closed
  timelike curves},'' \href{http://dx.doi.org/10.1103/PhysRevD.42.1915}{{\em
  Physical Review D} {\bf 42} (1990) no.~6, 1915--1930}.

\end{thebibliography}\endgroup
\end{document}